# GEOLAB: GEOMETRY-BASED TRACTOGRAPHY PARCELLATION OF SUPERFICIAL WHITE MATTER

*Nabil Vindas[1], Nicole Labra Avila[1,4], Fan Zhang[2], Tengfei Xue[2,3], Lauren J. O'Donnell[2], Jean-François Mangin[1]*

[1]Université Paris-Saclay, CEA, Neurospin, France
[2]Harvard Medical School, USA
[3]University of Sydney, Australia
[4]University College London (UCL), United Kingdom

## ABSTRACT

Superficial white matter (SWM) has been less studied than long-range connections despite being of interest to clinical research, and few tractography parcellation methods have been adapted to SWM. Here, we propose an efficient geometry-based parcellation method (GeoLab) that allows high-performance segmentation of hundreds of short white matter bundles from a subject. This method has been designed for the SWM atlas of EBRAINS European infrastructure, which is composed of 657 bundles. The atlas projection relies on the precomputed statistics of six bundle-specific geometrical properties of atlas streamlines. In the spirit of RecoBundles, a global and local streamline-based registration (SBR) is used to align the subject to the atlas space. Then, the streamlines are labeled taking into account the six geometrical parameters describing the similarity to the streamlines in the model bundle. Compared to other state-of-the-art methods, GeoLab allows the extraction of more bundles with a higher number of streamlines.

## 1. INTRODUCTION

Short-range white matter connections include a set of fiber bundles that circumvent cortical folds while remaining adjacent to the cortical mantle, allowing the transfer of information between neighboring regions of the brain. However, they have been less studied than long-range connections despite their high interest for clinical research [9]. To study short and long-range connections, diffusion magnetic resonance imaging is used to produce a tractogram, which gives a representation of the brain's white matter pathways and is used to study their organization in a non-invasive manner [1]. Nonetheless, tractography algorithms generate millions of streamlines (pseudofibers) with a great proportion of spurious trajectories [16]. Hence, tractogram parcellation is needed to sort streamlines and organize them into bundles.

Tractography parcellation methods can be manual or automatic. The manual process consists of using a visualization tool to select the streamlines and building bundles based on expert anatomical definition. This cumbersome method requires a neuroanatomical expert because the results deeply depend on the user. Hence the need for automated methods which can be divided into three groups: region of interest (ROI) based, clustering-based and image-based. ROI-based algorithms use a parcellation of the brain to select streamlines connecting the regions of interest [22,26,27]. In clustering-based methods, an atlas of reference bundles (clusters) is used to label the tractogram streamlines according to their similarity with the atlas [6-8,24]. These approaches can be computationally expensive and strongly depend on the alignment with the atlas. For example, RecoBundles [6] was developed to efficiently extract deep white matter (DWM) bundles using a global and local registration in the streamline-space before tractography parcellation.

In recent years, several deep-learning-based methods have shown to reduce computation time while providing good quality parcellation [2,23,24]. For example, the recent image-based method TractSeg [23] uses a CNN to learn the presence and orientation of bundles in each voxel and then performs a probabilistic bundle-specific tracking using this information. With regard to cluster-based SWM parcellation, SupWMA [24,25] uses a PointNet inspired architecture to work directly in streamline space. This is combined with supervised contrastive learning to solve the challenges of SWM parcellation by creating a good representation of streamlines where SWM bundles can be easily identified and outliers can be efficiently filtered out.

Most current parcellation methods focus on DWM and are not adapted for the extraction of short-range connections. Therefore, in this paper, we propose a method called GeoLab inspired by RecoBundles [6] but addressing the challenges of SWM tractography parcellation. In the spirit of RecoBundles, we use a global and local streamline-based registration (SBR) [5] to align the tractogram to the atlas space, but we refine this strategy to overcome frequent ambiguities when identifying short bundles. For this purpose, a statistical analysis of the atlas is done to extract six bundle-specific geometric properties of the streamlines. Hence, these properties constitute the basis for the definition of the SWM-dedicated similarity measure used in our method.

## 2. METHODOLOGY

### 2.1. Datasets and preprocessing

One short bundle atlas [12] with 657 SWM bundles built from the HCP database [3] was used as a reference for SWM tractography-based parcellation. This Extended Short Bundles Atlas (ESBA), which is proposed by the European EBRAINS infrastructure, was built from subjects registered to the MNI space using a non-linear sulcus-driven registration (DISCO/DARTEL [13]) imposing the most stable sulci as anatomical constraints. The atlas bundles result from a top-down divisive hierarchical clustering strategy with a new similarity measure: the Minimum of the Maximum Euclidean distance after alignment (MMEA), which aligns the streamlines at their medial points and then computes their minimum average direct-flip (MDF) distance [4]. Compared to other state-of-the-art SWM atlases [7,8,15,28], the ESBA is strongly linked with cortical folding and it is the most comprehensive short-bundle atlas publicly available today. As this is an unique atlas, the tools available in the community are not adapted for the new challenges it poses. First, as the atlas was built using a similarity measure focusing on shape, there is ambiguity between the bundles with several of them having similar geometric properties that cannot be disentangled with the classic MDF distance. Second, the classic registration methods are not adapted to the ESBA as it was built using a sulcus-driven registration. Hence, an alignment taking into account the shape of the bundles is needed. Moreover, as the bundles are small structures they do not constitute big anatomical landmarks. Therefore, a registration method taking into account the neighboring structures is needed to avoid spurious alignment of the atlas bundles.

For experimental evaluation on semi-ground truth (SGT), we used the work done in [12] with 76 subjects of the ARCHI dataset [17] aligned to MNI space using the sulcus-driven transforms mentioned above. For each subject, a streamline-regularized deterministic tractography was used to produce a tractogram. Then, the tractogram's size was reduced using an intra-subject clustering method proposed by [14], producing a tractogram of centroids representing the geometry of the streamlines in the raw tractogram. We obtained the final tractogram, called mean tractogram, by fusing the reduced tractogram of all the subjects. To label the centroids in this dataset, we used a semi-automatic iterative process. For each bundle in the atlas, we used a bundle-specific threshold on the MDF to label the centroids with similar trajectories to the atlas bundles. For a specific bundle, we chose the threshold by trying different values and selecting the one giving the best visual result. We managed to extract a total of 591 bundles, composed by centroids of the mean tractogram. The remaining 66 did not have a threshold giving satisfying results after visual check. This was expected as the resolution of ARCHI dataset is lower than the HCP dataset, hence not all geometric trajectories in the ESBA are present in ARCHI mean tractrogram.

For experimental evaluation without ground truth, we randomly selected 100 subjects (age: 48 to 78) from the UKBiobank dataset (https://www.ukbiobank.ac.uk/). The diffusion parameters are as follows : b =1000 and 2000 s/mm$^2$ with 50 directions each, resolution = 2,0 mm$^3$. We used the FSL's flirt tool to compute an affine transform between the diffusion and the structural (T1) data. Then, we computed linear and non-linear transforms from T1 to MNI space with FSL's flirt tool and ANTs toolbox. We processed the diffusion MRI data using MRtrix [21]. We used the multi-shell multi-tissue constrained spherical deconvolution model [10] with Anatomical-Constrained Tractography [18] and a second-order integration over the fiber orientation distribution [20] to produce 10 million streamlines. Then, we used the spherical-deconvolution Informed Filtering of Tractogram [19] to filter and reduce the tractography to 5 million streamlines. Finally, we registered the result to the MNI152 space using the transforms mentioned above. This registration serves as an initialization for the streamline-based alignment described below.

## 2.2. Geometrical labeling

### 2.2.1. Computation of geometric bundle descriptors

We performed an analysis of the ESBA using 6 parameters (figure 1) describing the geometrical features of the bundles. The motivation behind using geometrical features comes from the use of a shape-based similarity measure (MMEA) for the atlas inference [12]. First, we considered the length of the streamlines, the Euclidean distance of the middle points of streamlines to the barycenter of the bundle and the pairwise MMEA distance [12] between streamlines. Then, three novel features were considered: the angle between planes, the angle between directions and the shape angle. For each streamline, a plane was fitted, and the normal vector to this plane was used to compute the angle between the planes of two streamlines. Next, using the two vectors given by the middle point and the two extremities of the streamline, we defined a direction vector as the mean of these two vectors, which allows us to compute the angle between directions. Finally, we defined the shape angle as the angle formed between the two vectors used to compute the direction angle. This angle describes how close to a U-form a streamline is. We deliberately chose to define simple geometrical parameters to make our method computationally efficient. This allowed us to do a novel statistical analysis of the atlas to address the ambiguity found in SWM by fitting a bundle-specific distribution of the geometrical features. For each parameter the burr, log-normal, gamma, beta and normal distributions were tested, and the best distribution was chosen based on the sum of squares error. Finally, bundle-specific thresholds for each parameter were defined using the first and last decile.

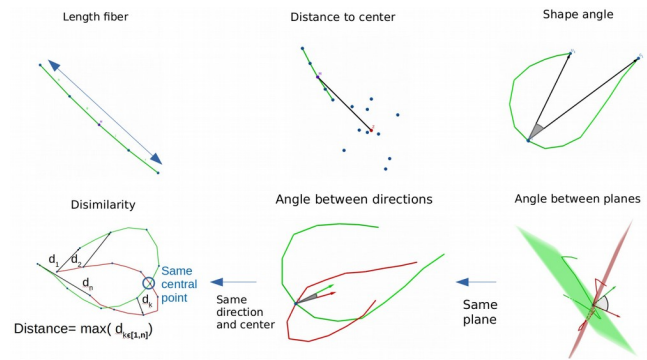

**Figure 1** : Geometric parameters used for fiber labeling : length of fiber, distance to barycenter of bundle, angle between planes, angle between direction, shape angle and disimilarity.

### 2.2.2. Local streamline neighborhood registration

When dealing with the ESBA, because of the numerous ambiguities across the bundles, it is crucial to use a local streamline neighborhood registration (LSNR). Using the classic approach to match the subject centroids to the atlas as in [6] might lead to mistakes because each bundle would lead to several hits. To solve this problem, we defined the radius of a bundle ($r_b$) as the radius of a sphere centered at the barycenter ($c_b$) of the bundle containing all its streamlines. We called the neighborhood of a bundle in a tractogram the streamlines of the tractogram contained in a sphere centered at $c_b$, with a radius of 6 times $r_b$. Thus, for each bundle we extracted a neighborhood in the whole atlas ($A_n$) and in the aligned tractogram (S) and we computed the centroids of each neighborhood ($A_n$' and S') using QuickBundles [4]. The proposed LSNR then registers S' to $A_n$' using the SBR method [5], and the resulting rigid transform is applied to the S.

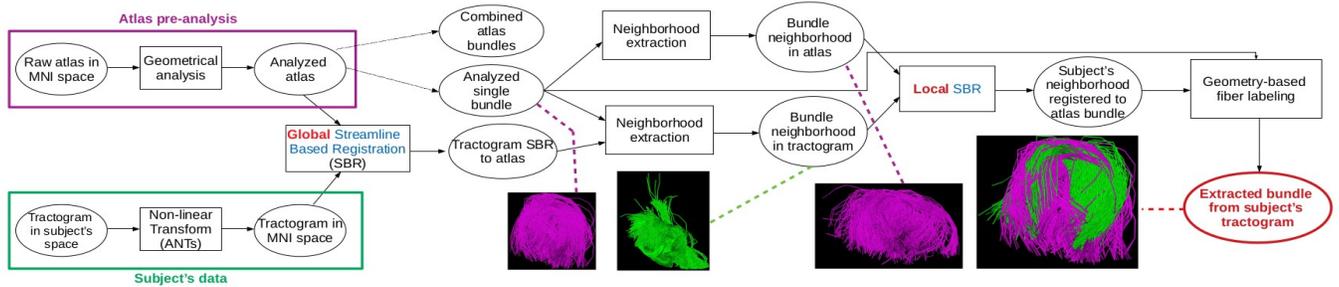

**Figure 2**: Projection pipeline for atlas-based tractography segmentation with three main steps : atlas analysis, streamline-based registration and geometric-based fiber labeling.

*2.2.3. Parcellation based on the 6 geometric features*

Figure 2 shows an overview of the method proposed here. After a standard prealignment of a subject's tractogram to the MNI152 space, a global streamline-based registration (SBR) [5] is done to refine the prealignment. Then, for each bundle a LSNR is done. Finally, using geometrical thresholds provided by the atlas statistical analysis (Section 2.2.1), we labeled the streamline according to their similarity to the model bundles using the geometrical features defined above. We tested different combinations of these features and the best results were obtained when using all 6 of them.

## 3. EXPERIMENTS AND RESULTS

### 3.1. Performance on semi-ground truth

We evaluated the performance of our method using the SGT, and we compared it to two state-of-the-art methods. First, RecoBundles with a threshold for the MDF of 5mm. Second, SupWMA, which has shown promising results for SWM parcellation. SupWMA is trained using a dataset containing labeled DWM and SWM bundles and outliers. As our atlas contains only short streamlines, we did not include DWM outliers. Moreover, to create a category of SWM outliers, we applied to the tractogram used to build the atlas the same semi-automatic process used to create the SGT, and the outliers were defined as the streamlines without any label. The hyperparameters for SupWMA were the same as in [24]. As the SGT was already in MNI space, no image-based registration was necessary.

**Table 1**: Performance scores on the semi-ground truth tractogram (mean (median) ± standard deviation * 100)

| Method | Sensitivity | Precision | Jaccard | F1-score |
|---|---|---|---|---|
| GeoLab | 72(81)±29 | 52(56)±31 | 42(45)±24 | 55(62)±31 |
| RecoBundles | 11(2)±16 | 2(0)±5 | 2(0)±3 | 3(0)±5 |
| SupWMA | 26(17)±24 | 36(36)±25 | 15(14)±12 | 24(25)±18 |

As in [6], the scores used to evaluate the performance of the models are the sensitivity, the specificity, the accuracy, the precision, the Jaccard index, the f1-score, the bundle adjacency (BA), the number of streamline per bundle (NFPB), and the percentage of bundles extracted (PBE) out of the 591 bundles in the SGT. The BA is a recent metric defined by [4] that gives the percentage of streamlines with neighbors between two bundles based on a threshold over the MDF distance (5 mm in our study). All the models got a specificity and accuracy of 1.0. Indeed, the size of a bundle is negligible relative to the size of the tractogram hence, the number of true negatives will be much larger than the number of false positives and false negatives [6]. Table 1 shows that GeoLab got the best overall scores but the precision, Jaccard index and F1-score were low for all models. Similar to the explanation given in [6] for RecoBundles, this can be explained by our method producing bundles with more centroids than the SGT bundles. This is expected as the MDF, used to build the SGT, does not allow to catch all centroids of a short-bundle because variability due to position and shape are mixed [12]. However, as seen in table 2, SupWMA did not produce bundles with more centroids than the bundles in the SGT, meaning that the model is filtering out labeled streamline. An explanation for this could be the SWM outliers used for the training and the geometrical ambiguity of the atlas as several outliers have a geometrical trajectory similar to labeled streamline. For RecoBundles, the very low scores suggest a large number of incorrect labels.

**Table 2**: BA, number of streamlines per bundle (SPB) recognized (SPB Rec) and PBE (a bundle is considered as valid if it has a minimum of 1 fiber for PBE-1 and 10 fibers for PBE-10) (mean (median) ± standard deviation, BA is shown in percentage).

| Method | BA to atlas | BA to SGT | SPB Rec | PBE-1 | PBE-10 |
|---|---|---|---|---|---|
| GeoLab | 85(90)±15 | 94(97)±9 | 157(113)±149 | 100 | 98 |
| RecoBundles | 75(77)±10 | 65(68)±13 | 1034(948)±632 | 99 | 99 |
| SupWMA | 88(90)±12 | 82(88)±18 | 55(35)±62 | 98 | 81 |

Table 2 shows a better BA to the SGT for the bundles extracted by GeoLab, indicating a higher similarity to the SGT than for the other methods. However, SupWMA has a better BA to the atlas, meaning that it is extracting bundles with higher similarity to the atlas as its filtering process is stringent. Moreover, after a visual check of the extracted bundles for all the methods, we found a possible explanation for the better performance of GeoLab. For SupWMA, there is great ambiguity in our atlas, with multiple cases of two geometrically similar bundles overlapping, or outliers having similar geometrical trajectories as the labeled centroids, and we do not have enough data to allow the network to build a latent space that resolves this ambiguity. In the case of RecoBundles, the use of the MDF is problematic as it does not allow to resolve this ambiguity because it combines in one measure several geometrical properties of the streamlines. Moreover, RecoBundles' use of the reference bundle during the SBR instead of a neighborhood can produce a misalignment of the reference bundle to neighboring structures with similar geometric properties. In other words, even if the similarity to the atlas is high, the misalignment produce labeled fibers with different labels than the SGT.

## 3.2 Performance on dataset without ground truth

We performed experiments on the UKBiobank dataset to evaluate the performance of our method in unseen data. Moreover, the age range of the 100 subjects from the UKBiobank dataset is [48-78] whereas the atlas was built with subjects between 22 and 36 years old. Finally, the tractography algorithm used to produce the atlas differs from the one used to produce the tractograms for the UKBiobank subjects, which allows us to test the robustness of our method to the tractography algorithm used. To evaluate the quality of the extraction, we used the overlap, coverage and bundle adjacency as defined in [4], and the PBE. The overlap gives the average number of neighbors in the model bundle for a streamline in the extracted bundle. The coverage gives the percentage of streamlines of the recognized bundle with neighbors in the model bundle. Hence, a high coverage and overlap indicates that a labeled streamline has a high similarity to the atlas.

**Table 3** : Mean for 100 UKBiobank subjects of BA to atlas, coverage, overlap, SPB recognized (SPB Rec) and PBE-1 as in table 2.

| Method | BA | Coverage | Overlap | PBE-1 | SPB Rec |
|---|---|---|---|---|---|
| **GeoLab** | 0.65$\pm$0.03 | 0.76$\pm$0.03 | 32$\pm$3 | 74$\pm$2 | 226$\pm$43 |
| **RecoBundles** | 0.69$\pm$0.12 | 0.57$\pm$0.10 | 18$\pm$3 | 90$\pm$17 | 2375$\pm$413 |
| **SupWMA** | 0.64$\pm$0.04 | 0.84$\pm$0.03 | 31$\pm$3 | 66$\pm$6 | 147$\pm$32 |

Table 3 shows that our method managed to extract around 74 % of the bundles in the atlas with a high similarity to the reference bundles. RecoBundles gave the highest BA but a lower coverage and overlap, meaning that most streamlines in the extracted bundle have neighbors in the reference bundle but the recognized bundle did not approximate the atlas as well as the other methods. Moreover, RecoBundles had a higher PBE-1, however a visual check showed that several of the recognized bundles are composed of spurious streamlines coming from the local SBR using the atlas bundle instead of a neighborhood. GeoLab seems like a good compromise between SupWMA and RecoBundles, having a quality of extraction similar to SupWMA but allowing recognition of more bundles with a higher number of streamlines. Figure 3 shows three different scenarios of extracted bundles for a random subject. While our method and SupWMA could produce higher quality bundles than RecoBundles (a), in some cases our method produced bundles with more streamlines and some spurious streamlines whereas SupWMA and RecoBundles did not show a good extraction (b). The bundle in (b) is a case of a bundle with high similarity to other bundles in the atlas, showing the ambiguity in our atlas. SupWMA and RecoBundles had a problem resolving this ambiguity. Finally, GeoLab was able to catch bundles that SupWMA could not and for which RecoBundles showed poor performance (c).

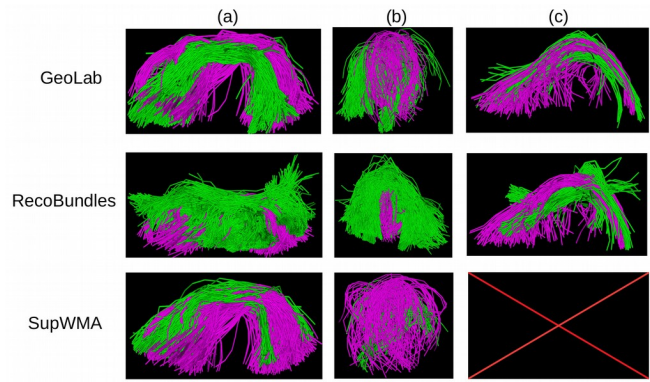

**Figure 3** : Visualisation of 3 recognized bundles for a random subject from the UKBiobank dataset for GeoLab, RecoBundles and SupWMA. IN purple the model bundle and in green the extracted bundle.

## 5. CONCLUSION

GeoLab is a parcellation algorithm adapted to the geometry of superficial white matter with good performance on ground truth and unseen data. It allows the segmentation of short-bundles with a high similarity to the reference bundles. On a ground truth dataset, comparison to other state-of-the-art methods showed that GeoLab allows to resolve the problem of the geometrical ambiguity found in our short-bundle atlas and it is a good compromise between quality of extraction and number of bundles recognized. However, compared to SupWMA, it does not include an outlier detection step which can cause outliers to be included in the extracted bundles. Moreover, our method requires a lot of memory making it unadapted for computers with low resources. Some improvement could be made by giving more importance to the two endpoints of streamlines as [25] suggests this information is important for SWM classification. Finally, the memory problem comes from the precomputation of the neighborhoods, which could be solved by computing them on the fly. GeoLab's codes and ESBA atlas are openly available at https://github.com/vindasna/GeoLab

## 7. ACKNOWLEDGEMENTS

This project has received funding from the European Union's Horizon 2020 Research and Innovation Programme under Grant

Agreement No. 945539 (HBP SGA3). We acknowledge the following NIH grant : R01MH125860